\documentstyle[aps,prl,epsfig]{revtex}

\begin{document}

\draft

\twocolumn[\hsize\textwidth\columnwidth\hsize\csname@twocolumnfalse\endcsname

\title{		Fluctuations of entropy production in driven glasses
}

\author{	Mauro Sellitto\cite{Present}
}

\address{
		Dipartimento di Scienze Fisiche and Unit\`a INFM,\\
 Universit\`a ``Federico II", Mostra d'Oltremare,
		Pad.~19, I-80125 Napoli, Italy
}

\date\today

\maketitle

\begin{abstract}
We study by Montecarlo simulation the non-equilibrium stationary behavior
of a three-dimensional stochastic lattice gas with {\it reversible\/}
kinetic constraints and in diffusive contact with two particle reservoirs
at {\it different\/} chemical potential.
When one of the boundaries is placed in the ``glassy phase'', the statistics
of entropy production fluctuations driven by the chemical potential gradient
satisfy the Gallavotti-Cohen fluctuation theorem provided that an
``effective'' entropy production is introduced.
\end{abstract}


\twocolumn\vskip.5pc]\narrowtext

\paragraph*{Introduction.}

The Gallavotti-Cohen fluctuation theorem (FT) is a result of remarkable
generality
concerning the distribution of entropy production fluctuations
over long time intervals in stationary open systems.
It states that the ratio of the probabilites of observing a given
average entropy production $\sigma_{\tau}$, over a time interval $\tau$,
to that of observing the opposite value $-\sigma_{\tau}$ is
${\rm e}^{\tau \sigma_{\tau}}$~\cite{GaCo}.
It has been recently shown \cite{Ga} that this result implies the
fluctuation-dissipation theorem (FDT) and Onsager's reciprocity
relations in the limit of vanishing driving forces (i.e.,
in equilibrium).
Therefore, the FT appears as an important stepping stone towards the
generalization of thermodynamics to situations out of equilibrium.

The FT was inspired by a pioneering numerical experiment on a thermostated
fluid under shear flow~\cite{EvCoMo}.
It has been further confirmed by the numerical simulation of an electrical
conduction model~\cite{BoGaGa}, and heat diffusion in a Fermi-Pasta-Ulam
chain~\cite{LeLiPo}, while a turbulent Rayleigh-B\'enard convection
experiment is still under analysis~\cite{CiLa}.
Although the theorem  was originally proved for thermostated Hamiltonian
systems driven by external forces, under certain ``chaoticity'' assumptions
for the dynamics~\cite{GaCo}, it also holds for rather ``generic'' stochastic
systems endowed with Langevin~\cite{Ku}, and diffusive dynamics~\cite{LeSp}.

It is well known, however, that violations of FDT appear in a wide  class
of conservative systems, since their relaxational dynamics 
after a quench in the low-temperature or high-density phase
is so slow that prevents them from reaching equilibrium on finite time
scales~\cite{BoCuKuMe}.
In the attempt to formulate a thermology of aging systems, the FDT violation
has been related to the appearance of an {\it effective temperature},
$T_{\rm eff}=T/X$, different from that of the thermal bath, $T$~\cite{CuKuPe}
(for a different point of view, see~\cite{Ni}).
The factor $X$ is called the fluctuation-dissipation (FD) ratio: it is
equal to one at high temperatures (where FDT holds), while it may become
smaller than one for sufficiently long times and at low enough temperatures.
In view of the connection between FDT and FT we can ask ourselves what form
assumes the latter in those driven systems for which a FDT violation
is expected in the limit of small entropy production.
We have therefore considered the following question.
Given a conservative aging system, ${\cal C}$, what is the form of FT for
the corresponding dissipative system, ${\cal D}$, obtained by putting
${\cal C}$ in contact with two reservoirs (one of which in the glassy phase)?
The results of our computer experiment suggest that for such a driven
system a generalized form of FT holds where the entropy production
is evaluated taking into account the FD ratio, $X$, of the corresponding
conservative, undriven, system.

\paragraph*{The driven glass.}
To corroborate this conjecture we consider a three dimensional lattice-gas
model of hard-core particles kept away from equilibrium by a chemical
potential gradient (see fig.~\ref{box}).
At the top and bottom layer of the system particles are inserted and
removed according the usual Montecarlo rule, mimicking the contact of
the system with a particle reservoir ${\cal R}_{\pm}$:
we randomly choose a site on the layer; if it is empty, we add a new
particle; otherwise we remove the particle with probability
${e}^{-\beta \mu_{\pm}}$ ($\mu_{\pm}\ge 0$).
The global effect of the reservoirs is to fix the boundary densities
at two different values $\rho_+$ and $\rho_-$, driving a current through
the system.
A similar setup was previously consider by Spohn~\cite{Sp} with the aim of
studying analytically long-range correlations in a simple model of
non-equilibrium fluid subject to temperature gradients.
The link of the microscopic theory with the fluctuating hydrodynamics
was also investigated, see~\cite{EyLeSp1} and, for a review,~\cite{Sp_book}.
Here, we consider a bulk dynamics implementing the cage effect in supercooled
liquids~\cite{KoAn}.
Therefore, the sweeps of creation/destruction of particles on the edges
are alternated with the following diffusive sweep.
A particle and one of its neighbouring sites are chosen at random;
the particle moves if the three following conditions are all met:
1) the neighbouring site is empty;
2) the particle has less than $4$ nearest neighbours;
3) the particle will have less than $4$ nearest neighbours after
it has moved.
The rule is symmetric in time and, unlike the usual driven
diffusive systems where detailed balance is broken~\cite{ScZi,EyLeSp},
the system is microscopically reversible.

The dynamical behavior of the system  when the two reservoirs are at same
chemical potential has been investigated in~\cite{KuPeSe,PeSe,Se}.
It is worth to recall briefly some results.
Although the equilibrium thermodynamics of the model is trivial,
a purely dynamical transition takes place when the system is
quenched in the glassy phase, namely above the threshold value
$\mu_{\rm c} \simeq 2.0$:
the density approaches by a power-law an asymptotic value
$\rho_{\rm c} \simeq 0.88$,
different from the equilibrium value allowed by the dynamical evolution rule,
and the diffusion coefficient of the particles vanishes roughly as the
inverse of the time elapsed after the quench~\cite{KuPeSe,PeSe}.
Consequently the system ages, and the mean-square displacement and the
conjugated response function satisfy a generalized FDT with a FD ratio 
given by $X \simeq 0.79$~\cite{Se}.

We are interested here to the case where a boundary is in the fluid phase 
($\mu_- < \mu_{\rm c}$) and the other one in the glassy phase
($\mu_+ > \mu_{\rm c}$).
The numerical experiment is performed in the following way.
The system consists of a cubic lattice of size $L^{2} \times 2L$
with $L=10$. The reservoir  ${\cal R}_+$ is located at $z=0$ while
${\cal R}_-$ at $z=\pm L$. In this way we can take periodic boundary
conditions in all directions and avoid spurious edge effects~\cite{Mu}.
We start with a low-density uniform distribution of particles
and let the system reach a stationary state characterized by
a time-translation invariant correlation function of current
fluctuations.
We then follow the dynamical trajectory of the motion for a time of
$10^8$ Montecarlo sweeps (MCs) along which the observables of interest 
are evaluated.
In particular we consider the particle current $J(t)$, defined as an
extensive quantity (proportional to the transverse surface $L^2$) by:
\begin{eqnarray}
	J(t)
& = &
	\frac{1}{2L} \sum_{z=-L}^{L-1}
	\left[ j_+(z,t)-j_-(z,t) \right] ,
\end{eqnarray}
where $j_+(z,t)$ and $j_-(z,t)$  are the number of jumps taking place
at time $t$ through the layer $z$ in the direction of, and opposite to, 
respectively, the externally imposed chemical potential gradient.

\paragraph*{Fluctuation theorem and effective entropy production.}
In order to study the statistics of current fluctuations we consider
the average particle current over a time interval of duration $\tau$:
\begin{eqnarray}
J_{\tau}(t) & = & \frac{1}{\tau} \sum_{s=t+1}^{t+\tau} J(s) ,
\end{eqnarray}
which in a steady state does not depend on $t$.
We then compute the probability distribution $\pi_{\tau}(p)$ of the
adimensional variable
\begin{eqnarray}
	p
& = &
	\frac{J_{\tau}}{J} \,
\end{eqnarray}
where $J = \lim_{\tau \rightarrow \infty} J_{\tau}$.
The function  $\pi_{\tau}(p)$ also gives the distribution of entropy
production fluctuations since, according the thermodynamics of linear
irreversible processes~\cite{Pr}, the average entropy production
$\sigma_{\tau}$ over a time interval $\tau$ is simply related to the
corresponding average particle current $J_{\tau}$ by:
\begin{eqnarray}
\sigma_{\tau}
	& = &
J_{\tau} \,( \mu_+-\mu_-) \,.
\end{eqnarray}
Fig.~\ref{p} shows the distribution $\pi_{\tau}(p)$ for different values
of the time interval, $\tau=1,2,5,10$ MCs.
The chemical potential of the reservoirs is $\mu_+=2.2$ and $\mu_-=0$.
Being the system in the ``large deviation'' regime it can appear 
surprising that the shape of $\pi_{\tau}(p)$, is compatible with a 
gaussian distribution, even though no {\it a priori\/} relation is 
expected with the Sinai limit theorem~\cite{Si}.
For a discussion see~\cite{BoGaGa}, where a similar result was also obtained;
while a skew distribution was found in~\cite{LeLiPo}.

Once determined the probability distribution $\pi_{\tau}(p)$ of
entropy-production
fluctuations it is simple to check the Gallavotti-Cohen FT,
that in our case reads:
\begin{eqnarray}
\log\frac{\pi_{\tau}(p)}{\pi_{\tau}(-p)}
	& = &
\tau \, p \, J \, (\mu_+-\mu_-) .
\label{GC}
\end{eqnarray}
Fig.~\ref{ft} shows that the lhs and rhs of the relation~(\ref{GC})
are linearly related but are not equal.
However, an equality is obtained by taking into account 
the ``effective chemical potential'' $\mu_{\rm eff}= X \mu$
(or, equivalently, the FD ratio $X$) of the 
glassy boundary in the calculation of the entropy production:
\begin{eqnarray}
\sigma_{\tau}
	& = &
J_{\tau} \,( \mu_{\rm eff}-\mu_-) \,.
\end{eqnarray}
When the reservoirs are both placed in the fluid phase, where $X=1$ (and 
$\mu_{\rm eff}=\mu_+$), the FT should be recovered.
The inset of fig.~\ref{ft} shows, consistently, that for $\mu_+=1.8$,
$\mu_-=0$, and $\tau=1,2,5$ MCs the FT is well satisfied.

To rule out that these results are a mere coincidence we have also
performed extensive numerical simulations with different values of 
the chemical potential $\mu_-$.
In fig.~\ref{ft_eff}, we show the results for 
$\mu_-=-0.4,-0.2, 0.0, 0.2, 0.4$, and $\mu_+=2.2$, 
with $\tau=1$ MCs, plotted against the 
effective entropy production.
We see that also this plot supports the conjecture of a generalized
form of FT.

We now give an heuristic argument showing how the appearance of the FD ratio
here should not be surprising.
Let us observe that our experimental setup is nothing else than a device
to perform measurements of the effective chemical potential, $\mu_{\rm eff}$,
of the system~\cite{nota1}.
Indeed, suppose that we want to measure the value of $\mu_{\rm eff}$
for an aging system in contact with a reservoir ${\cal R}_+$ at
$\mu_+>\mu_{\rm c}$.
We start with the stationary system in contact with ${\cal R}_+$ and
${\cal R}_-$, with $\mu_-<\mu_{\rm c}$, the last reservoir acting as a
``thermometer''.
Then, if we disconnect the system from ${\cal R}_-$ and let it relax,
the density of the corresponding layer will increase until the average
particle current is zero (the value of $\mu$ where this condition is
first met defines $\mu_{\rm eff}$).
However, since the system is unable to fully equilibrate with the
reservoir ${\cal R}_+$ (being $\mu_+>\mu_{\rm c}$), such a density
will be lower than $\rho_{\rm c}$ and therefore the corresponding
chemical potential will be reduced by a factor $X$, i.e.,
$\mu_{\rm eff}=X \, \mu_+$.
In principle, this method represent a possible way to measure the
FD ratio of aging systems and gives some insight into the notion of
effective temperature as partial equilibration factor~\cite{CuKuPe}.

It is also interesting to observe that the stationary density profile
can be predicted by a non-linear diffusion equation~\cite{PeSe}.
However, this matter as well as the validity of Einstein relation between
diffusivity and conductivity will be discussed elsewhere.

\paragraph*{Conclusions and perspectives.}
We have investigated  by Montecarlo simulation non-equilibrium stationary
behavior of a constrained lattice-gas model driven by a chemical potential
gradient (driven glass).
The approach does not involve fictitious thermostating mechanism allowing
dissipation to prevent the heating up of the system~\cite{NoHo},
neither the question of the interpretation of phase space contraction
rate as entropy production, which has not yet received an unanimous
answer (see however~\cite{Ru1}).
The extension to other systems such as driven spin-glasses and to different
boundary conditions modelling, {\it e.g.}, a Couette flow, is also
feasible~\cite{Fi}.

In particular, we have studied the distribution of entropy production
fluctuations when one of the reservoirs is placed in the glassy phase,
$\mu_+>\mu_{\rm c}$, and shown that in the stationary state a generalized
form of FT is satisfied which takes into account the effective entropy
production computed through the FD ratio of the corresponding undriven 
system.
Therefore, the generalized form of FT and FDT should be related by
the same limiting procedure exploited in~\cite{Ga}.
This vindicates the crucial role played by the notion
of ``effective temperature'' in generic, aging or stationary, 
out of equilibrium systems.

% characterized by a small entropy production.

Although the model considered here is much simpler than a real supercooled
liquid in a temperature gradient, it displays interesting phenomena that
could be observed, hopefully, also in a ``true'' experiment.

We conclude by observing that it is not clear what happens in a generic
driven glass in which the reservoirs are both placed in the glassy phase,
at different values of temperature.
In such a strongly non-linear regime of small entropy production the
correlation of current fluctuations could exhibit aging and the system
never reach a stationary state.
What would be the form of Gallavotti-Cohen FT in such a situation,
is presently unknown.

%\acknowledgements

The author thanks L. Peliti for a critical reading of the manuscript
and F. Bonetto, G. Gallavotti and H. Spohn for discussions.
Partial support from INFM through the contract 1229/UDR-NAPO-AMM 
is acknowledged.

%\break

%%%%%%%%%%%%%%%% figure %%%%%%%%%%%%%%%%%%%%%
\begin{figure}[f]
\begin{center}
%\vspace*{5.45cm}
\hfill \epsfig{file=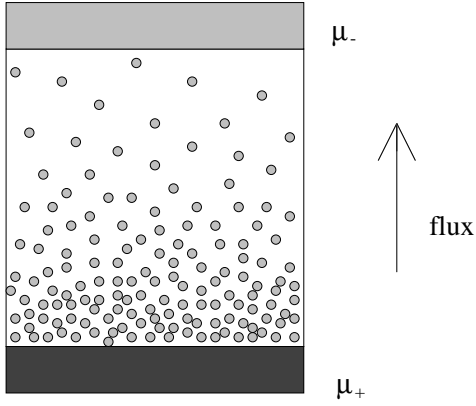,width=5.25cm,angle=270}
%\vspace{-4.5cm}
\end{center}
\caption{Rayleigh-B\'enard experiment to study
out of equilibrium stationary states in driven glasses.
The system is coupled to two particle reservoir at different
chemical potential, the lower one in the glassy phase, $\mu_+>\mu_{\rm c}$,
and the upper one in the fluid phase, $\mu_-<\mu_{\rm c}$.
Particles flow through the bulk from the bottom to the top.}
\label{box}
\end{figure}
%%%%%%%%%%%%% end of figure %%%%%%%%%%%%%%%%%

%%%%%%%%%%%%%%%% figure %%%%%%%%%%%%%%%%%%%%%
\begin{figure}[f]
\begin{center}
%\vspace{1cm}
\epsfig{file=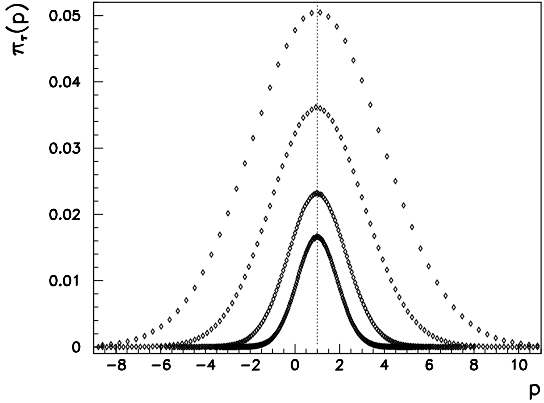,width=8.5cm}
%\vspace{1cm}
\end{center}
\caption{Histograms of the probability distribution $\pi_{\tau}(p)$
of entropy production fluctuations along a trajectory of motion of
$10^8$ MCs sampled at time intervals $\tau=1,2,5,10$ MCs
(from top to bottom).
The chemical potential on the boundaries is $\mu_+=2.2$ (glassy phase)
and $\mu_-=0$ (fluid phase).
The curves are all consistent with a gaussian distribution
of average 1.}
\label{p}
\end{figure}
%%%%%%%%%%%%% end of figure %%%%%%%%%%%%%%%%%

%%%%%%%%%%%%%%%% figure %%%%%%%%%%%%%%%%%%%%%
\begin{figure}[f]
\begin{center}
%\vspace{1cm}
\epsfig{file=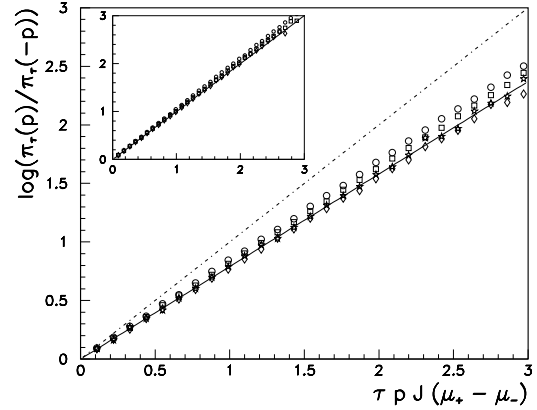,width=8.5cm}
%\vspace{1cm}
\end{center}
\caption{ 
Logarithmic probability ratio $\log (\pi_{\tau}(p)/\pi_{\tau}(-p))$
vs. the entropy production over a time interval $\tau$, for
$\tau=1$ (diamonds), 2 (stars), 5 (squares), 10 (circles) MCs;
and $\mu_+=2.2$, $\mu_-=0$.
The broken line, with slope $1$, represents the prediction of the FT.
The straight line, with slope $X$, take into account the effective chemical
potential $\mu_{\rm eff}=X \mu_+$ of the glassy boundary.
The value of FD ratio, $X=0.79$, is taken from ref.~\protect\cite{Se}.
Inset: check of FT when both the reservoirs are placed in the fluid 
phase, $\mu_+=1.8$ and $\mu_-=0$, for $\tau=1$ (diamonds), 
$2$ (squares), $5$ (circles) MCs.  
}
\label{ft}
\end{figure}
%%%%%%%%%%%%% end of figure %%%%%%%%%%%%%%%%%

%%%%%%%%%%%%%%%% figure %%%%%%%%%%%%%%%%%%%%%
\begin{figure}[f]
\begin{center}
%\vspace{1cm}
\epsfig{file=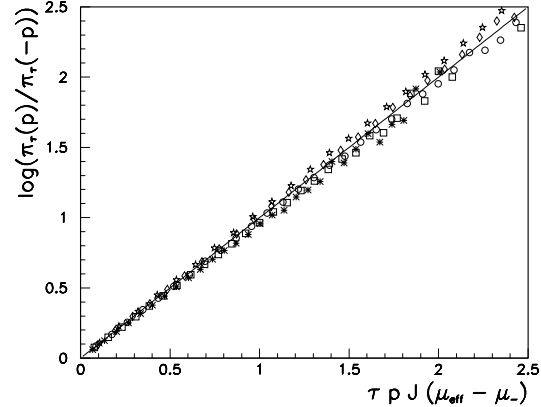,width=8.5cm}
%\vspace{1cm}
\end{center}
\caption{%
Logarithmic probability ratio vs. the effective
entropy production for $\tau=1$ MCs, $\mu_+=2.2$, and 
$\mu_-=-0.4$ (asterisks), $-0.2$ (squares), 0 (circles), 0.2 (diamonds), 
0.4 (stars).
The straight line with slope 1 represents the prediction of generalized FT,
where $\mu_{\rm eff}=X \mu_+$ and the value of FD ratio, $X=0.79$, 
is taken from ref.~\protect\cite{Se}.}
\label{ft_eff}
\end{figure}
%%%%%%%%%%%%% end of figure %%%%%%%%%%%%%%%%%


\begin{references}
\bibitem{nota1} To pursue better the analogy with usual thermal system
		consider that particle density, particle current and
		chemical potential play the same role of energy density,
		heat flux, and inverse temperature, respectively.
\end{references}
\end{document}